**Nonreciprocal control and cooling of phonon modes in an optomechanical system**

H. Xu[1], Luyao Jiang[1], A. A. Clerk[2], J. G. E. Harris[1,3]

[1] Department of Physics, Yale University, New Haven, CT, 06520, USA

[2] Institute for Molecular Engineering, University of Chicago, Chicago, IL, 60637, USA

[3] Department of Applied Physics, Yale University, New Haven, CT, 06520, USA

**Phononic resonators play important roles in settings that range from gravitational wave detectors to cellular telephones. They serve as high-performance transducers, sensors, and filters by offering low dissipation, tunable coupling to diverse physical systems, and compatibility with a wide range of frequencies, materials, and fabrication processes. Systems of phononic resonators typically obey reciprocity, which ensures that the phonon transmission coefficient between any two resonators is independent of the direction of transmission[1,2]. Reciprocity must be broken to realize devices (such as isolators and circulators) that provide one-way propagation of acoustic energy between resonators. Such devices are crucial for protecting active elements, mitigating noise, and operating full-duplex transceivers. To date, nonreciprocal phononic devices[3,4,5,6,7,8,9,10,11] have not combined the features necessary for robust operation: strong nonreciprocity, *in situ* tunability, compact integration, and continuous operation. Furthermore, they have been applied only to coherent signals (rather than fluctuations or noise), and have been realized exclusively in travelling-wave systems (rather than resonators). Here we describe a cavity optomechanical scheme that produces robust nonreciprocal coupling between phononic resonators. This scheme provides ~ 30 dB of isolation and can be tuned *in situ* simply via the phases of the drive tones applied to the cavity. In addition, by directly monitoring the resonators' dynamics we show that this nonreciprocity can be used to control thermal fluctuations, and that this control represents a new resource for cooling phononic resonators.**

Reciprocity is a generic feature of linear, time-invariant oscillator systems. It may be broken in various ways; for example by introducing bias, nonlinearity, or parametric time dependence[1,2]. In phononic systems, nonreciprocal bias can be introduced by imposing rotational motion[9] or a magnetic field[3,4,5]. However the former is impractical in many settings, and the latter typically produces weak nonreciprocity. Likewise, nonlinearity-based approaches[5,7,8] have required bulky components and generally result in signal distortion. In contrast, parametric modulation can produce nonreciprocity with considerable flexibility (as demonstrated recently in the photonic and microwave domains[12,13,14,15]). Parametric modulation of phononic resonators arises naturally in cavity optomechanical systems, which consist of an electromagnetic cavity that is detuned by the motion of mechanical oscillators[16]. In particular, electromagnetic drive tones applied to the cavity can tune the mechanical oscillators' frequencies, dampings, and couplings, an effect known as "dynamical backaction"[16]. This effect has been used to realize transient nonreciprocity (by adding a slow time dependence to the parametric modulation[10,11]); in contrast, the scheme described here uses stationary modulation and achieves continuous operation.

The phononic resonators studied here are two normal modes of a SiN membrane[17] with dimensions 1 mm × 1 mm × 50 nm. We focus on a pair of low-order drumhead-like modes with resonant frequencies $\omega_1 = 2\pi \times 557.473$ kHz and $\omega_2 = 2\pi \times 705.164$ kHz and damping rates $\gamma_1 = 2\pi \times 0.39$ Hz and $\gamma_2 = 2\pi \times 0.38$ Hz. The membrane is positioned inside a cryogenic Fabry-Perot optical cavity with linewidth $\kappa = 2\pi \times 180$ kHz and coupling rate $\kappa_{in} = 2\pi \times 70$ kHz (for light with wavelength $\lambda = 1,064$ nm). The mechanical resonators couple to the cavity with rates $g_1 = 2\pi \times 2.11$ Hz and $g_2 = 2\pi \times 2.12$ Hz. The device's construction and characterization are described in Refs. [10,11]. The wide separation between $\omega_1$ and $\omega_2$ allows the motion of both modes to be inferred from a single record of the cavity detuning, which is provided by a probe laser that drives the cavity with fixed intensity and detuning.

Near-resonant coupling can be induced between these modes by modulating the dynamical backaction at a frequency close to $\delta\omega \equiv \omega_1 - \omega_2$. Such modulation arises from the intracavity beat note produced when the cavity is driven by two tones whose detunings (relative to the cavity resonance) are $\Delta_1 = -\omega_1 + \Delta_\ell$ and $\Delta_2 = -\omega_2 + \Delta_\ell$.[18,11,19] As illustrated in Fig. 1a, the optomechanical interaction allows a photon to scatter from one drive tone to the other by transferring a phonon between the modes. This process occurs via a virtual state in which the photon is at a mechanical sideband of the drive tones. The participation of the various mechanical sidebands can be enhanced or suppressed by the cavity's resonance; for the detunings shown in Fig. 1a, the cavity ensures that the sideband with detuning $\Delta_\ell$ is the dominant path by which phonon transfer takes place.

This phonon transfer process has two crucial features. First, the transfer amplitude involves the complex-valued cavity susceptibility $\chi(\Delta_\ell)$ (where $\chi(\omega) = (\kappa/2 - i\omega)^{-1}$) regardless of the direction of transfer, and so has both a dissipative and a coherent character. Second, the phase of the intracavity beatnote appears explicitly in the transfer coefficient. These features alone do not result in nonreciprocal energy transfer (for example, the beatnote phase can be gauged away). However as discussed in Refs. [20,21,22,12], interference between two such processes can break reciprocity.

Figure 1b shows the arrangement used for the experiments described here. The cavity is driven by four tones (these address a different cavity mode than the probe laser[10,11]). Their detunings $\Delta_{1,2,3,4}$ are chosen to provide two beat notes that each induce near-resonant coupling between the modes (i.e., $\Delta_1 - \Delta_2 = \Delta_3 - \Delta_4 \approx \delta\omega$) and hence two distinct copies of the phonon transfer process illustrated in Fig. 1a. The $\Delta_{1,2,3,4}$ are also chosen so that the dominant mechanical sideband in each transfer process has a distinct detuning: $\Delta_\ell = \Delta_2 + \omega_2 \approx \Delta_1 + \omega_1$ and $\Delta_u = \Delta_4 + \omega_2 \approx \Delta_3 + \omega_1$. As described below, interference between these two processes results in nonreciprocal energy transfer between the phononic modes. Moreover, this interference is controlled by the relative phase between the two beatnotes (which cannot be gauged away).

This system can be described via the standard linearized optomechanical equations of motion for one cavity mode and two mechanical modes[16,23] (see Methods). The cavity mode is subject to a drive of the form $\sum_{n=1}^{4} \sqrt{P_n} e^{i(\Delta_n t + \phi_n)}$ where $P_n$ is the power of the $n^{\text{th}}$ tone. The detuning, power, and phase ($\phi_n$) of each tone is set by a microwave generator (MWG) that produces the four tones from a single laser via an acousto-optic modulator. Adiabatically eliminating the cavity field leaves equations of motion for the two mechanical mode amplitudes that correspond to the effective time-dependent Hamiltonian (see Methods):

$$H = \begin{pmatrix} \omega_1 - i\gamma_1/2 + f_1 & [g e^{i(\theta_\ell + \phi_{12})} + h e^{i(\theta_u + \phi_{34})}] e^{i\Delta t} \\ [g^* e^{i(\theta_\ell - \phi_{12})} + h^* e^{i(\theta_u - \phi_{34})}] e^{-i\Delta t} & \omega_2 - i\gamma_2/2 + f_2 \end{pmatrix}$$

The diagonal elements of $H$ represent the usual single-tone dynamical backaction: $f_a \approx \sum_{n=1}^{4} P_n g_a^2 |\chi(\Delta_n)|^2 \chi(\Delta_n + \omega_a)$ where $a \in \{1,2\}$. In contrast, the off-diagonal components of $H$

describe the coupling between the two mechanical modes mediated by the intracavity beatnotes. The phases of these beatnotes are $\phi_{12} \equiv \phi_1 - \phi_2$ and $\phi_{34} \equiv \phi_3 - \phi_4$. The coefficients $g \approx \sqrt{P_1 P_2} g_1 g_2 \chi^*(\Delta_1)\chi(\Delta_2)|\chi(\Delta_\ell)|$ and $h \approx \sqrt{P_3 P_4} g_1 g_2 \chi^*(\Delta_3)\chi(\Delta_4)|\chi(\Delta_u)|$. For clarity, the present discussion ignores smaller terms in *f, g,* and *h* that are due to non-resonant mechanical sidebands (these terms are included in the analysis and fits presented below, and in the full description in Methods). Lastly, $\theta_{\ell,u} \equiv \arg(-i\chi(\Delta_{\ell,u}))$ and $\Delta \equiv \Delta_1 - \Delta_2$.

Isolation between the two mechanical modes (e.g., corresponding to $H_{1,2} = 0 \neq H_{2,1}$) can be achieved by first choosing $P_n$ and $\Delta_n$ so that $|g| = |h|$. For the present device, this is realized with all the $P_n$ = 5 μW and $\Delta_n = \{-\omega_1 + \Delta_\ell + \zeta, -\omega_2 + \Delta_\ell, -\omega_1 + \Delta_u + \zeta, -\omega_2 + \Delta_u\}$ where $\Delta_\ell = -2\pi \times 60$ kHz and $\Delta_u = 2\pi \times 150$ kHz. The constant $\zeta$ is the detuning of the beat notes relative to $\delta\omega$, and is set to $2\pi \times 100$ Hz. With the condition $|g| = |h|$ satisfied, $\phi_{12}$ and $\phi_{34}$ may be adjusted via the MWG to ensure that one off-diagonal element of *H* vanishes while the other does not. This is shown in Fig. 1c, which plots $H_{1,2}$ and $H_{2,1}$ as a function of $\phi \equiv \phi_{12} - \phi_{34}$. For $\phi \approx \pi/2$, *H* allows energy to flow from mode 1 to mode 2 but not *vice versa*. The situation is reversed when $\phi \approx -\pi/2$. In contrast, $\phi \approx 0$ gives $H_{1,2} \approx H_{2,1}$. This tunability between isolation, reciprocity, and reversed isolation occurs while keeping the $P_n$ and $\Delta_n$ fixed, and only varying the $\phi_n$. This avoids cross-talk between the nonreciprocity and other device parameters (such as the mechanical frequencies, which depend only weakly on the $\phi_n$).

To demonstrate the nonreciprocity's tunability, we measured the transfer of energy between the two modes for various choices of $\phi$. Two measurements with $\phi = \pi/2$ are shown in Fig. 1d, which plots $\varepsilon_1(t)$ and $\varepsilon_2(t)$: the energy in each mode (as inferred from the probe beam) as a function of time *t*. For *t* < 0 the control tones are off, and one mode is driven to an average energy ~ $10^{-18}$ J (corresponding to amplitude ~ $5 \times 10^{-11}$ m). The other mode is undriven, except by thermal fluctuations consistent with the bath temperature $T_{\text{bath}}$ = 4.2 K. At *t* = 0 the drive is turned off and the control tones are turned on for a duration $\tau$ = 3 ms. For *t* > $\tau$ the control tones are off again. Fig. 1d demonstrates the isolation described above: under the influence of control tones with $\phi = \pi/2$, an excitation prepared in mode 1 is transferred to mode 2 (upper panel) while an excitation prepared in mode 2 is not transferred to mode 1 (lower panel).

Figure 2 shows the energy transmission coefficients $T_\uparrow \equiv \varepsilon_2(\tau)/\varepsilon_1(0)$ and $T_\downarrow \equiv \varepsilon_1(\tau)/\varepsilon_2(0)$ (corresponding to transfer from mode 1 to mode 2 and *vice versa*) as a function of $\tau$ and $\phi$. The performance is summarized in Fig. 3, which shows the same data converted to the isolation $I \equiv T_\uparrow/T_\downarrow$. The damping of the modes ensures that $T_\uparrow$ and $T_\downarrow$ decrease with $\tau$ (Fig. 2); however $I$ is nearly independent of $\tau$ (Fig. 3). Figures 2 & 3 both show that reciprocity is restored (corresponding to $|H_{1,2}| = |H_{2,1}|$) for a slightly non-zero value of $\phi$, reflecting the fact that the nonreciprocity of $H$ is determined by the phases $\theta_\ell$ and $\theta_u$ as well as by $\phi$. The data in Fig. 3 show that this system achieves $I \geq 30$ dB from mode 1 to mode 2 and $I \leq -25$ dB in the opposite direction. It also shows that $I$ can be tuned over this entire range (including through 0 dB) by varying the $\phi_n$ while all other parameters are held fixed. The solid lines in Figs. 2 & 3 are not fits, but rather the time evolution predicted by the matrix exponential of $H$.

Experiments on non-reciprocal devices (in the phononic as well as other domains) typically measure the scattering matrix describing propagating waves incident on and emanating from the device. In contrast, the measurements described here provide direct access to the device's internal degrees of freedom. This opens the possibility of controlling the resonators' state via their nonreciprocal interactions. To demonstrate this, we use the nonreciprocity described above to modify the resonators' thermal fluctuations and to realize a form of cooling with no equivalent in reciprocal systems.

To describe the system's steady-state fluctuations, we note that both modes couple to the thermal bath ($T_{\text{bath}} = 4.2$ K) and to the cavity field (whose effective temperature can be approximated as zero for the present discussion[16]). In the absence of coupling between the phononic modes, these two "baths" would cause each mode to equilibrate to a temperature $T_a = (\gamma_a/\text{Im}[f_a])T_{\text{bath}}$ where $a \in \{1,2\}$ and we assume the single-tone optical damping $\text{Im}[f_a] \gg \gamma_a$. This reduction of $T_a$ with respect to $T_{\text{bath}}$ is the well-known effect of "cold damping" or "laser cooling".[16] However in the present system the modes also couple to each other. When the resulting energy transport is reciprocal ($|H_{1,2}| = |H_{2,1}|$) thermal phonons are exchanged between the modes, tending to bring $T_1$ and $T_2$ closer together. In contrast, if $H$ is chosen to give unidirectional energy transport (e.g., for $\phi = \pm\pi/2$), then the isolated mode emits thermal phonons into the other mode but not *vice versa*. This leads to cooling of the isolated mode and heating of the other mode, even if the former is initially the colder of the two.

To realize this isolation-based cooling we use the same $\Delta_n$ as above and $P_n = 2.5$ μW (resulting $H_{1,2}$ and $H_{2,1}$ as in Fig. 1c but reduced by a factor of two). No external drive is applied to the phononic modes, and their undriven motion is recorded by the probe laser. Fig. 4a shows the spectral density of each oscillator's energy $S_{E_1}$ and $S_{E_2}$ for $\phi = -\pi/2$, 0, and $+\pi/2$. For all values of $\phi$, the mechanical linewidth is dominated by Im[$f_a$] (which is independent of $\phi$). The asymmetric lineshapes in Fig. 4a result from interference between the two modes (this is less apparent in $S_{E_2}$ owing to its narrower linewidth). The solid lines in Fig. 4a are fits to the expected form (a constant background plus the square modulus of the sum of two Lorentzians).

To measure the impact of nonreciprocity on the mode temperatures, $T_1$ and $T_2$ are determined from the area under the peaks in $S_{E_1}$ and $S_{E_2}$ at several values of $\phi$ (as in Fig. 4a). The result is shown in Fig. 4b as the normalized temperature difference $\Theta(\phi) \equiv 1 - \left(T_1(\phi)/T_2(\phi)\right)/\langle T_1/T_2 \rangle$ where $\langle \ldots \rangle$ denotes the average over $\phi$. Maximizing the isolation between the modes (i.e., setting $\phi = \pm\pi/2$) results in the most extreme values of $\Theta$. We emphasize that changing the sign of $\Theta$ is equivalent to reversing the direction of heat flow between the modes. Since $\langle T_1/T_2 \rangle = 0.56$ in these measurements, the optomechanical coupling transports heat from the colder mode to the hotter mode when $\Theta > 0$.

The solid line in Fig. 4b shows $\Theta$ as calculated from the optomechanical equations of motion (Methods). The agreement between the measured and predicted cooling extends over a wide range of parameters, as illustrated in Fig. 4b-e which show $\Theta(\phi)$ for various $\Delta_n$. The main impact of varying $\Delta_n$ is to break the condition $|g| = |h|$, resulting in weakened isolation and suppression of $\Theta$.

We also emphasize that the data in each of Fig. 4b-e were taken with fixed $P_n$ and $\Delta_n$, and that the additional cooling of one mode is accomplished just by varying the control tones' phases. Since conventional laser cooling techniques (e.g., those using the single-tone dynamical backaction) are independent of these phases, this shows that the nonreciprocity demonstrated here represents an additional resource for controlling the thermal fluctuations of phononic resonators.

In conclusion, we have demonstrated a robust, compact, and tunable scheme for inducing nonreciprocal dynamics between phononic resonators. We have applied this nonreciprocal control to external signals as well as to the resonators' thermal motion. The nonreciprocity is produced by a cavity optomechanical interaction, but the same scheme can be realized in other oscillator systems with parametric controls, including those in the electrical, mechanical, and optical domains.[24,25,26]

**Figure Captions**

**Figure 1 | Optically induced mechanical nonreciprocity. a,b** Top panel: frequency-domain illustration of the optomechanical control scheme. Black curve: cavity lineshape. Thin coloured arrows: control tones. Thick coloured arrows: motional sidebands that dominate the phonon transfer process. Horizontal axis: detuning from cavity resonance. Bottom panel: energy-domain illustration of the same scheme. Solid horizontal lines: states labelled by the number of phonons in each mode and the number of cavity photons. Dashed horizontal lines: virtual states through which the transfer process occurs. Cavity linewidth is indicated by the grey shading. The absolute frequency of the $i^{\text{th}}$ control tone is indicated by $\Omega_i$. The scheme shown in **a** is for illustrative purposes, while **b** shows the scheme used in the present work. **c,** The off-diagonal elements of the effective Hamiltonian $H$. The real and imaginary parts of $H_{1,2}$ and $H_{2,1}$ (in units of Hz) are plotted parametrically versus the phase $\phi$ between the two beatnotes. The control beam powers and detunings are given in the main text. The colour scale encodes the value of $\phi$. The grey (brown) arrow indicates the evolution of $H_{1,2}$ ($H_{2,1}$) as $\phi$ is increased from 0. **d,** Measurement of the mechanical energy in each mode as a function of time. The control beams are on only during the grey region (0 ms $\leq t \leq$ 3 ms). Data for $t >$ 3 ms is fit to a decaying exponential (black curves), and this fit is extrapolated to $t =$ 3 ms to find $\varepsilon_{1,2}(\tau)$, the energy in each mode at the end of the control pulse (black dots). Identical control beams (with $\phi = \pi/2$) are used in both panels, but energy is transferred only from mode 1 to mode 2.

**Figure 2 | Nonreciprocal phonon transmission.** The energy transmission coefficients $T_\uparrow$ and $T_\downarrow$ as a function of the control tones' duration $\tau$ (first three panels) and phase $\phi$ (last panel). Each point is determined from measurements similar to those in Fig. 1d. The solid lines are the theoretical prediction described in the main text.

**Figure 3 | Isolation between phononic resonators.** The isolation ratio $I$ as a function of the control tones' duration $\tau$ and phase $\phi$. The values of $I$ are extracted from the data in Fig. 2. The solid lines are the theoretical prediction described in the main text.

**Figure 4 | Cooling by nonreciprocity. a,** The power spectral density of the two modes' thermal motion. For clarity, the data have been offset horizontally so that the two modes (which oscillate at 557 kHz and 705 kHz) can be directly compared. From left to right, the three panels correspond to $\phi = -\pi/2, 0, +\pi/2$. **b-e,** The normalized difference between the two modes' temperatures. In each panel, the control beam detunings are as given in the main text, plus an additional offset $\Delta_{\text{off}}$.

**Acknowledgements** This work is supported by AFOSR grant FA9550-15-1-0270, AFOSR MURI FA9550-15-1-0029, ONR MURI N00014-15-1-2761, and the Simons Foundation (Award Number 505450).

**Author contributions** H.X., A.A.C. and J.G.E.H. designed the study. H.X. and L.J. carried out the measurements. H.X. and L.J. analyzed the data. All authors contributed to the writing of the manuscript.

**Competing Interests** The authors declare no competing interests.

# Figure 1

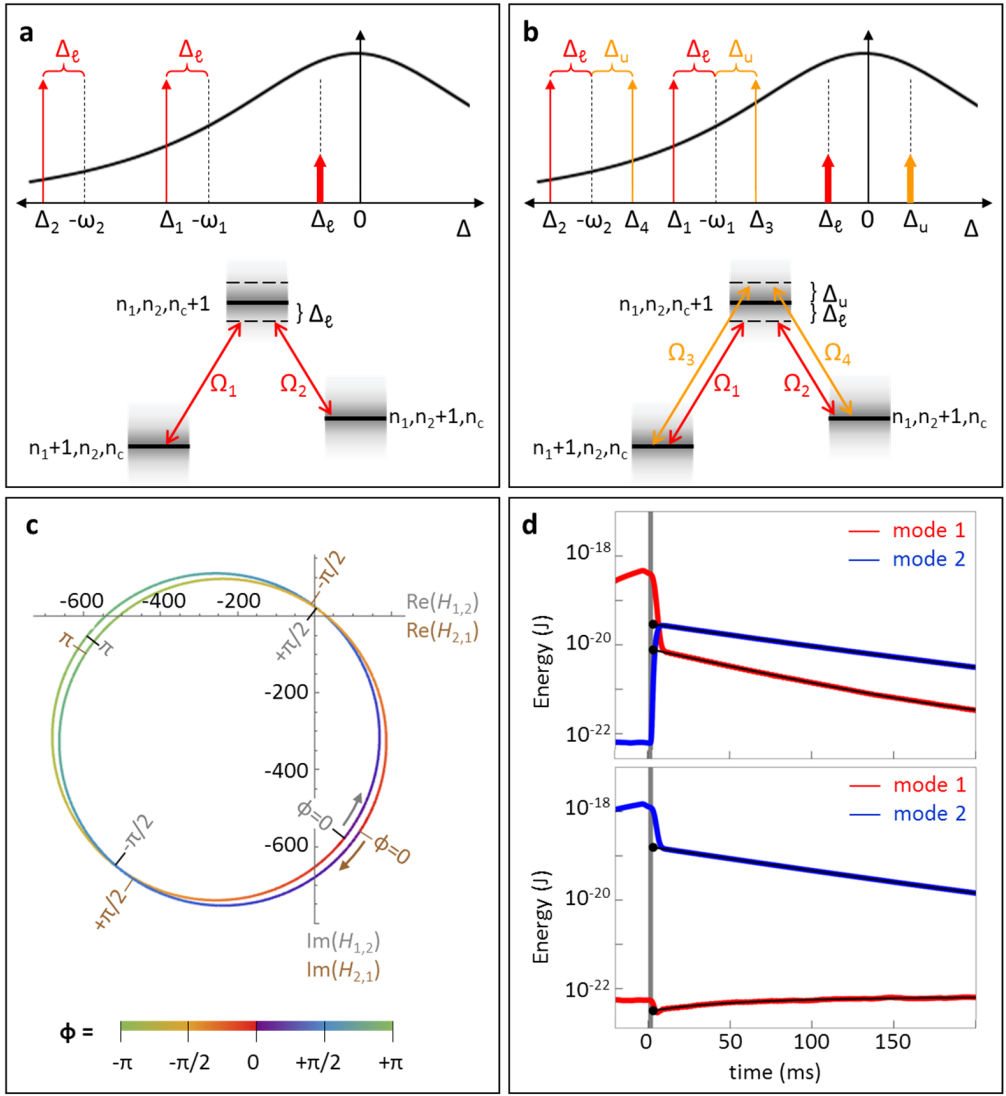

**Figure 2**

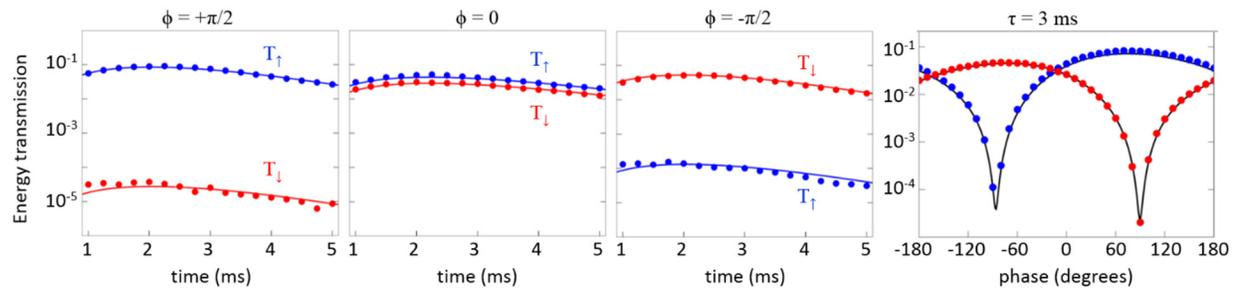

**Figure 3**

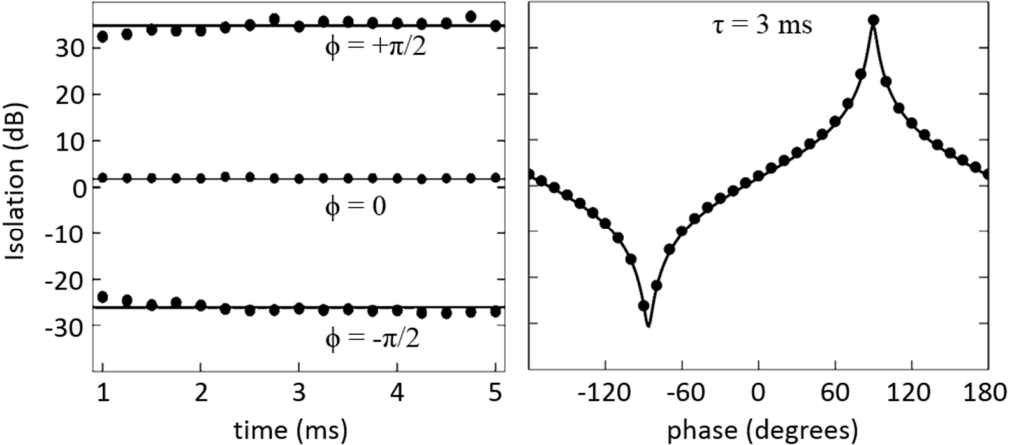

**Figure 4**

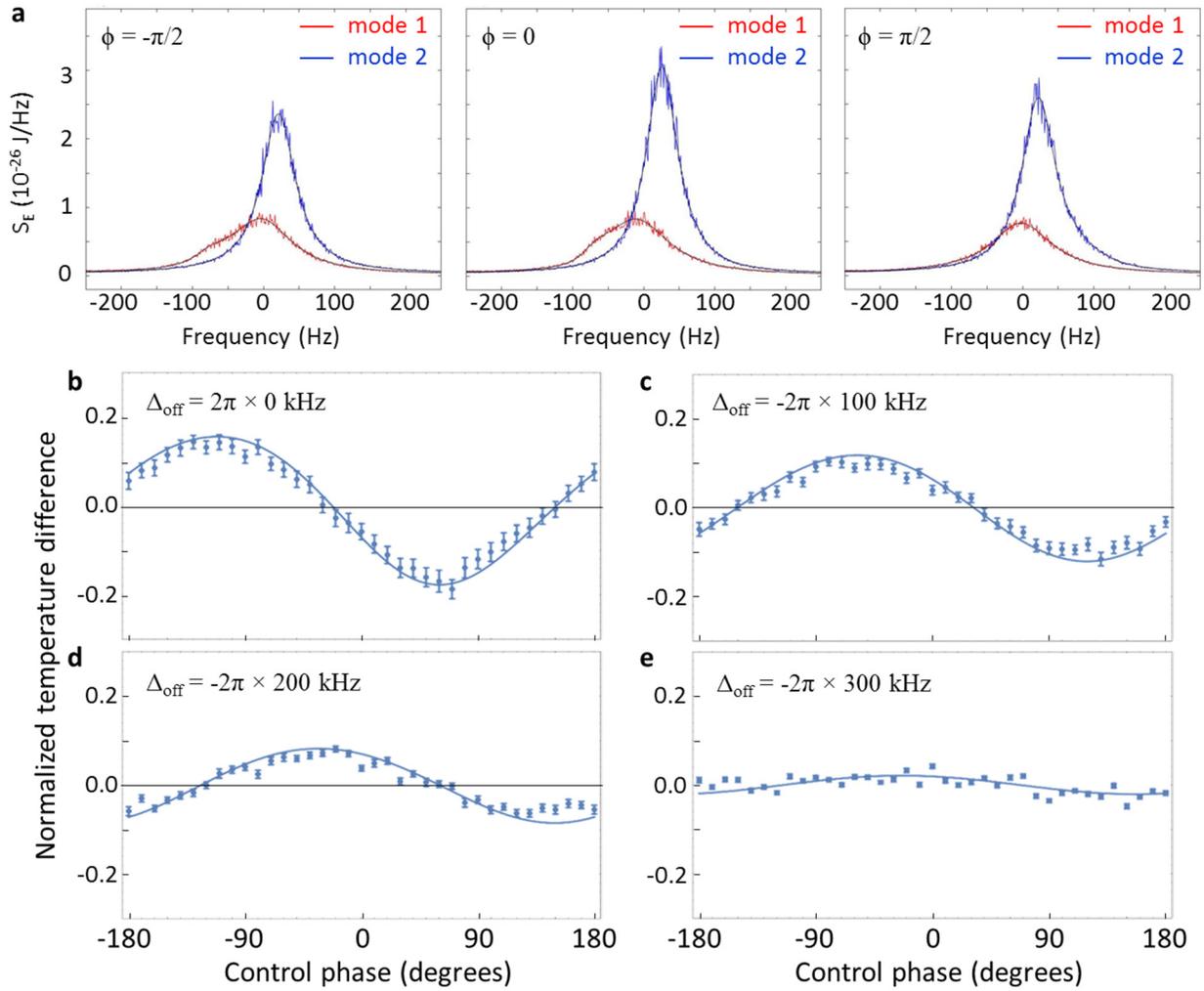

**METHODS**

**Theoretical model.** We consider two phonon modes coupled to a single optical mode via the usual optomechanical interaction described by the Hamiltonian $H_{\text{OM}} = \sum_{n=1}^{2} \hbar g_n (c_n + c_n^\dagger) a^\dagger a$. Here $\hbar$ is the reduced Planck's constant, $g_n$ is the single-photon coupling strength between the $n^{\text{th}}$ phonon mode and the optical mode, $a$ is the optical mode's annihilation operator, and $c_n$ is the annihilation operator for the $n^{\text{th}}$ phonon mode.[16,23] The equations of motion for the modes are then:

$$\dot{c}_1 = (-\frac{\gamma_1}{2} + i\omega_1) c_1 - i g_1 a^\dagger a + \sqrt{\gamma_1} \eta_1 \qquad (1)$$

$$\dot{c}_2 = (-\frac{\gamma_2}{2} + i\omega_2) c_2 - i g_2 a^\dagger a + \sqrt{\gamma_2} \eta_2 \qquad (2)$$

$$\dot{a} = (-\frac{\kappa}{2} + i\Omega_c) a - i(g_1(c_1 + c_1^\dagger) + g_2(c_2 + c_2^\dagger))a + \sqrt{\kappa_{\text{in}}} a_{\text{in}} \qquad (3)$$

where $\Omega_c$ is the cavity resonance frequency, and the $\eta_i$ and $a_{\text{in}}$ are the drives for, respectively, the phonon modes and the optical mode.

The cavity is driven by two pairs of control lasers to induce nonreciprocity between the phonon modes. The control lasers' detunings (with respect to the cavity resonance) are: $\Delta_1 = -\omega_1 + \Delta_\ell + \zeta$, $\Delta_2 = -\omega_2 + \Delta_\ell$, $\Delta_3 = -\omega_1 + \Delta_u + \zeta$, $\Delta_4 = -\omega_2 + \Delta_u$. Numerical values for these detunings are given in the main text (note that $\zeta \ll \omega_1, \omega_2, \Delta_\ell, \Delta_u$).

Stokes and anti-Stokes scattering of these control lasers can convert a phonon from one mechanical mode to the other. To describe this process quantitatively, we first linearize the optical field by the displacement $a = \alpha + d$, where $\alpha$ is the coherent amplitude of the optical mode and $d$ is the mode's fluctuations. The optomechanical interaction then becomes $H_{\text{OM,lin}} = \sum_{n=1}^{2} \hbar g_n (c_n + c_n^\dagger)(\alpha^* d + \alpha d^\dagger)$ where $\alpha = \sum_{k=1}^{4} \alpha_k e^{-i\Delta_k t}$ and $\alpha_k$ is the coherent amplitude contributed by the $k^{\text{th}}$ control laser.

The parameters of the experiment are such that the mechanical resonance frequencies and the separation between the motional sidebands are always much greater than the mechanical linewidths. As a result, the cavity field can be adiabatically eliminated to obtain the effective Hamiltonian for the mechanical modes:

$$H = \sum_{n=1}^{2} \hbar(\omega_n - i\gamma_n + \sigma_{n,n})c_n^\dagger c_n + \sigma_{1,2} e^{i\Delta t} c_1^\dagger c_2 + \sigma_{2,1} e^{-i\Delta t} c_2^\dagger c_1 \tag{4}$$

where

$$\sigma_{n,n} = \sum_{k=1}^{4} i\hbar g_n^2 |\alpha_k|^2 \left( \chi(\omega_n + \Delta_k) - \chi(\omega_n - \Delta_k) \right) \tag{5}$$

$$\sigma_{1,2} = \sum_{k=1}^{2} i\hbar g_1 g_2 \alpha_{2k-1}^* \alpha_{2k} e^{i(k-1)\phi} \left( \chi(\omega_1 - \Delta_{2k}) - \chi(\omega_1 + \Delta_{2k-1}) \right) \tag{6}$$

$$\sigma_{2,1} = \sum_{k=1}^{2} i\hbar g_1 g_2 \alpha_{2k-1} \alpha_{2k}^* e^{-i(k-1)\phi} \left( \chi(\omega_2 - \Delta_{2k-1}) - \chi(\omega_2 + \Delta_{2k}) \right) \tag{7}$$

and $\chi(\omega) = (\kappa/2 - i\omega)^{-1}$. As described in the main text, $\phi$ is the relative phase between the two beat notes whose frequencies are nearly equal to $\delta\omega$.